\documentclass[preprint,showpacs,preprintnumbers,amsmath,amssymb]{revtex4}

\newcommand{\bb}{\mathbf}

\usepackage{graphicx}
\usepackage{dcolumn}
\usepackage{bm}
\usepackage[T1]{fontenc}
\usepackage{hyperref}
\usepackage{setspace}
\usepackage[latin1]{inputenc}

\begin{document}

\title{Gate-controlled current switch in graphene}

\author{Kimmo S\"{a}\"{a}skilahti}
\author{Ari Harju}%
 \homepage{http://tfy.tkk.fi/qmp}
\affiliation{Helsinki Institute of Physics and Department of Applied Physics,\\
 Helsinki University of Technology, P.O.Box
4100, FI-02015 TKK, Finland}

\author{Pirjo Pasanen}
\homepage{http://research.nokia.com/}
\affiliation{Nokia Research Center, P.O.Box 407, FI-00045 Nokia Group, Finland}

\date{\today}

\begin{abstract}

We study numerically cross conductances in a four-terminal all-graphene setup. We show that far away from the Dirac point current flows along zigzag directions, giving the possibility to guide the current between terminals using a tunable pn-junction. The device operates as a gate-controlled current switch, and the electronic properties of graphene are crucial for efficient performance.

\end{abstract}

\maketitle

Graphene, the two-dimensional form of carbon, has been creating a lot of interest not only in the physics community but also in the electronic industry, due to its extraordinary physical properties \cite{Novoselov2004, Geim2007, Neto2009}. These properties include, for example, exceptionally high charge mobility, room temperature ballistic transport, ultrahigh thermal conductivity and mechanical strength. Provided that the manufacturing questions will be resolved, in the long term graphene could be used as material for high-performance nanoelectronic devices, even on flexible and transparent substrates. 



Keeping in mind the ultimate goal of building all-graphene circuits, it is important to study the electric properties of more complex geometrical structures. Multiterminal graphene devices have been studied theoretically in \cite{Laakso2008,Chen2008}. It was shown, for example, that near the Dirac point evanescent modes lead to quantum corrections in the multiterminal cross correlations \cite{Laakso2008}. Effects like these are interesting, since they probe the unique electronic properties of graphene. One of the most unusual consequences of the quasi-relativistic electron dynamics is the Klein tunneling \cite{Klein1929, Katsnelson2006, Beenakker2008}. The effect entails that electrons incoming to a potential barrier in graphene can couple to the hole states inside the barrier and transmit through perfectly, since backscattering is strictly forbidden in clean graphene pn-junctions \cite{Ando1998}. Theoretically, graphene nanodevices based on manipulating charge carriers by potential barriers have been proposed earlier in \cite{Cresti2008_2, Cheianov2007}. 

In this letter, we study multiterminal conductances in a four-terminal geometry and see how the intricate properties of graphene can be exploited even further. We show, for example, that the cross conductances between different terminals can be tuned by a step potential. In other words, the device performs as a current switch controlled by a top gate. 

\begin{figure}[t]
\includegraphics[width=8.5cm]{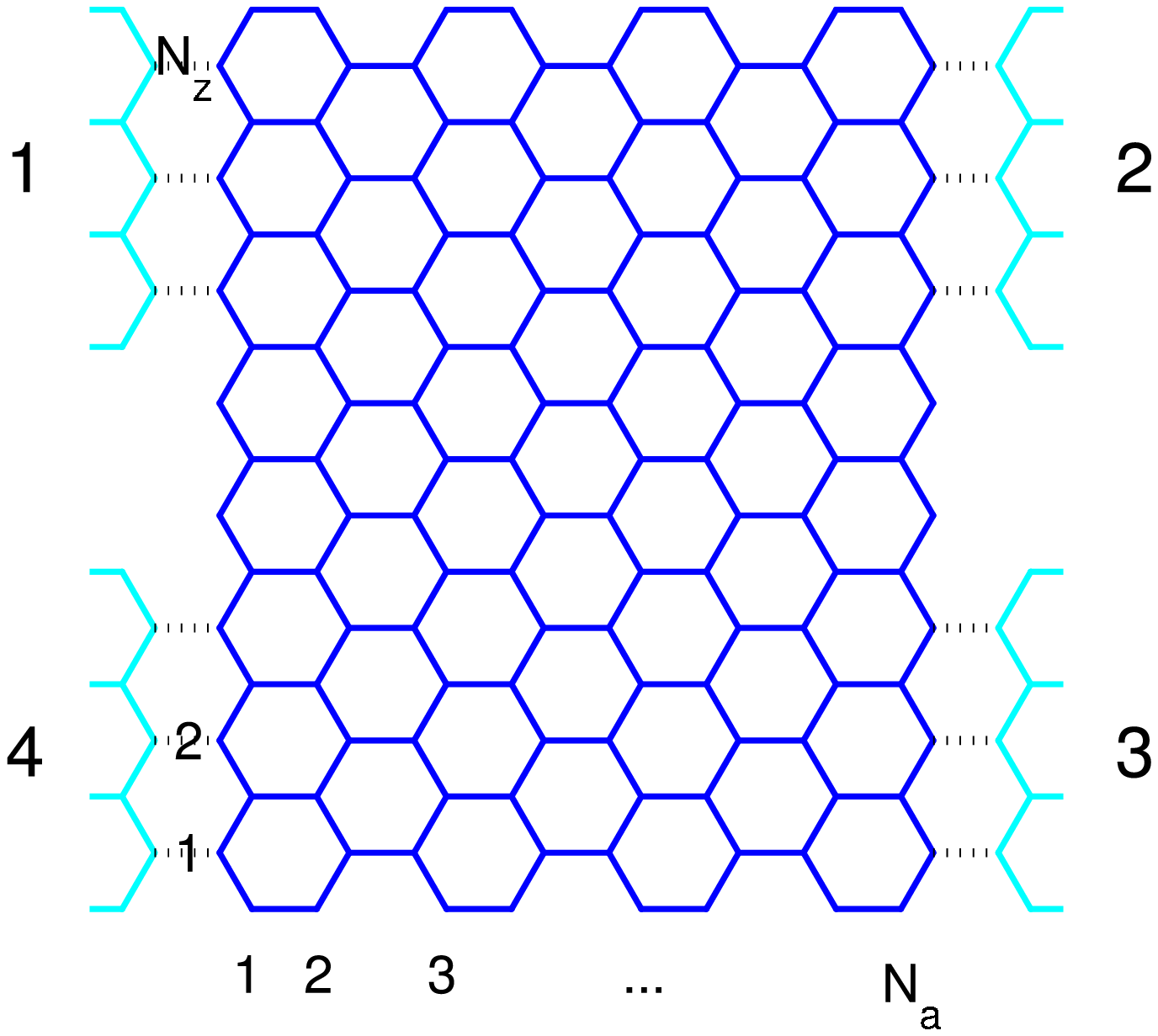}
\caption{(Color online) The setup for an $(N_a,N_z)$ device.}
\label{kuva1}
\end{figure}

The studied geometry is shown in Fig. \ref{kuva1}. The central region (device area) is a rectangular graphene island. The size of the the island is given by a pair of numbers $(N_a,N_z)$, where $N_a$ $(N_z)$ is the number of outermost atoms on the armchair (zigzag) edge, as shown in the figure. The width of the device is $W/a=\sqrt{3} N_z$ and the length $L/a=3N_a/2$, where $a \approx 0.142$ nm is the carbon-carbon distance. We attach left-right symmetric armchair leads onto the four corners of the island and study the cross conductances between the four terminals. One can view the first terminal acting as a source of electrons, and our aim is to control the current to one of the other leads. 

We calculate the currents numerically in the linear response and zero temperature regime, as described by Datta \cite{Datta1995}. 
The retarded Green's function for the device area is given by
 \begin{equation}
  G^R(E)=\left[(E+i 0^+)-H_D-\sum_{\alpha=1}^4 \Sigma_{\alpha}^R (E) \right]^{-1},
 \end{equation}
 where $H_D$ is the tight-binding Hamiltonian for the device area and $\{\Sigma_{\alpha}^R\}_{\alpha=1}^4$ are the self-energies describing the four semi-infinite leads. The energy $E$ can be interpreted as the overall Fermi level $E_F$ of the system. The tight-binding Hamiltonian of graphene is  $H = - \sum_{i,j} t_{ij} c_i^{\dagger} c_j$,
where $t_{ij}=t \approx 2.7$ eV for nearest neighbors and zero otherwise. 
The cross conductance between leads $\alpha$ and $\beta$ is given by the Landauer transmission formula 
\begin{equation}
G_{\alpha \beta} \equiv -\frac{dI_\alpha}{dV_\beta}= \frac{2e^2}{h} \textrm{Tr} \left[ \textbf{s}_{\alpha \beta}^{\dagger} \textbf{s}_{\alpha \beta} \right],
\end{equation}
where $\bb{s}_{\alpha \beta}$ is the scattering matrix describing electron transport from lead $\beta$ to lead $\alpha$ \cite{Buttiker1992}. The element $|s_{\alpha \beta}^{nm}|^2$ of the scattering matrix denotes the probability that a transversal mode $m$ in lead $\beta$ is scattered to mode $n$ in lead $\alpha$, and the trace $\textrm{Tr} \left[ \textbf{s}_{\alpha \beta}^{\dagger} \textbf{s}_{\alpha \beta} \right]$ is a sum over all these probabilities. 
The traces are calculated using the formula
\begin{equation}
G_{\alpha \beta}=\frac{2e^2}{h} \textrm{Tr} \left[\Gamma_\alpha G^R \Gamma_\beta \left(G^R\right)^{\dagger} \right],
\end{equation}
where the coupling matrices are given by the imaginary parts of the self-energies, i.e. $\Gamma_\alpha = i \left(\Sigma_\alpha^R-\left(\Sigma_\alpha^R\right)^{\dagger} \right)$.  Since we are working in linear response, we refer to ''current'' and ''conductance'' interchangeably. 

 \begin{figure}[t]
  \includegraphics[width=8.5cm]{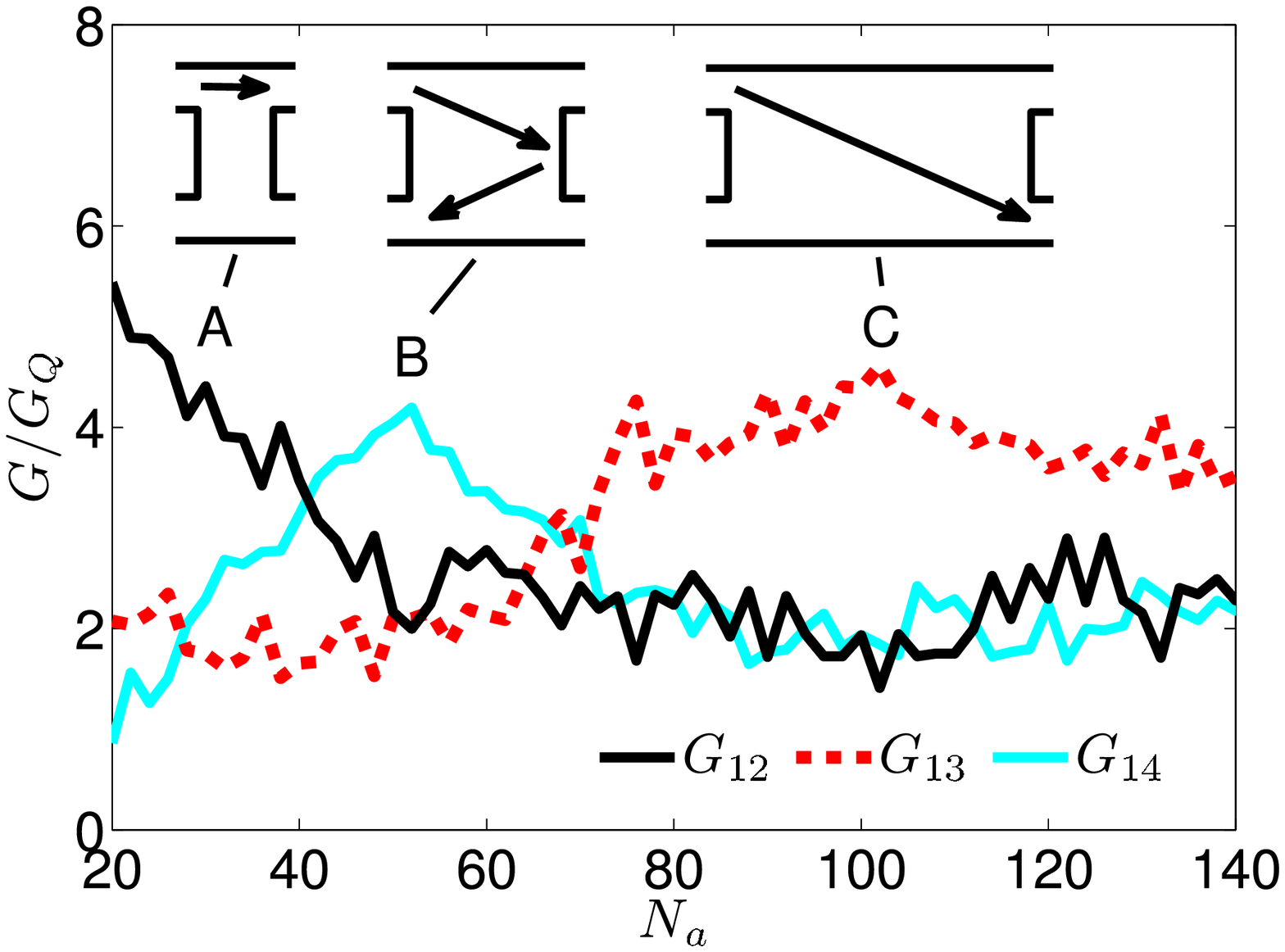}
  \caption{(Color online) The cross conductances as a function of the length $N_a$ for a device area with width $N_z=52$. Fermi level is at $E_F=0.8t$. Insets: Schematic representations of the current flow for selected aspect ratios (A-C).}
  \label{fig:length}
  \end{figure}

Let us start the analysis by looking at the effects of geometry. The width of device is fixed to be $N_z=52$ and we choose leads such that each lead is connected to $13$ atoms in the zigzag edges. For small Fermi energies, the current is nearly evenly distributed between all terminals (see below). However, for a high Fermi level one starts to see direction-dependent effects. 

The results for cross conductances at $E_F=0.8t$ as a function of the junction length $N_a$ are shown in Fig. \ref{fig:length}. The conductances are given in units of the conductance quantum $G_Q=2e^2/h$. There are nine propagating modes in the leads which is the maximum total transmission for leads of this width. Figure \ref{fig:length} shows many interesting features. In a short junction, the dominating cross conductance is $G_{12}$, meaning that most of the current travels straight from the terminal $1$ to the terminal $2$ (point A in the figure). Increasing the length makes the cross conductance $G_{12}$ smaller until at $N_a \sim 50$ the cross conductance $G_{14}$ peaks (point B). This can be interpreted as a reflection from the opposite wall, giving the first hint that the flow of current is not chaotic by nature.

When the length of the junction is still increased, the cross conductance $G_{13}$ starts to increase, peaking at $N_a \approx 100$ (point C). Interestingly, this corresponds to an aspect ratio of $\tan^{-1} (W/L)=30 ^{\circ}$. This proves that for this high a Fermi level, the current has a strong tendency to propagate along the zigzag path from terminal $1$ to terminal $3$. This is due to trigonal warping in the two non-equivalent corner points of the Brillouin zone \cite{Neto2009}.

\begin{figure}[t]
  \includegraphics[width=8.5cm]{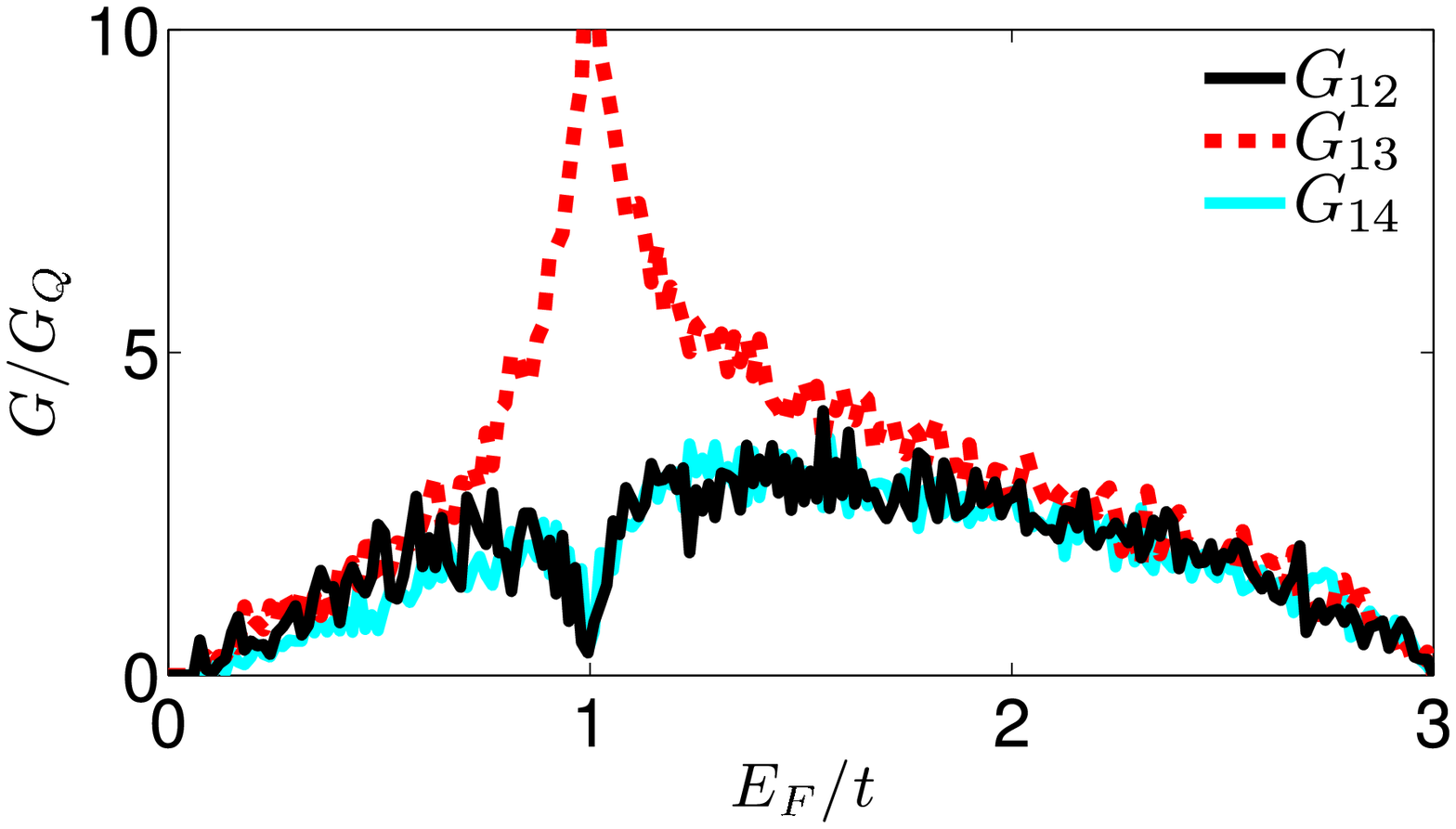}
  \caption{(Color online) The cross conductances as a function of the Fermi level for a $(104,52)$-junction. The zigzag direction is preferred only near $E_F = t$.}
  \label{fig:length2}
 \end{figure}

The requirement $L=\sqrt{3}W$ means roughly $N_a=2N_z$, so we fix the geometry to be $(104,52)$. The cross conductances as a function of the Fermi level for a $(104,52)$-junction are shown in Fig. \ref{fig:length2}. For $E_F \lesssim 0.5t$ cross conductances are nearly equal, indicating that all terminals are equally probable points for exit. Thus the behaviour of the currents inside the rectangular middle area could be described as chaotic. This is not surprising, since the relativistic dynamics in rectangular dots is known to be non-integrable \cite{Berry1987}. Only when $E_F \gtrsim 0.5t$ starts the cross conductance $G_{13}$ to dominate due to the tendency to propagate along the zigzag path. This tendency requires the Fermi level to be rather high, since in the Dirac cone all directions are on equal footing, and to get out of this regime the Fermi wavelength must be reduced to be of the same order as the lattice constant.

\begin{figure}[t]
  \includegraphics[width=8.5cm]{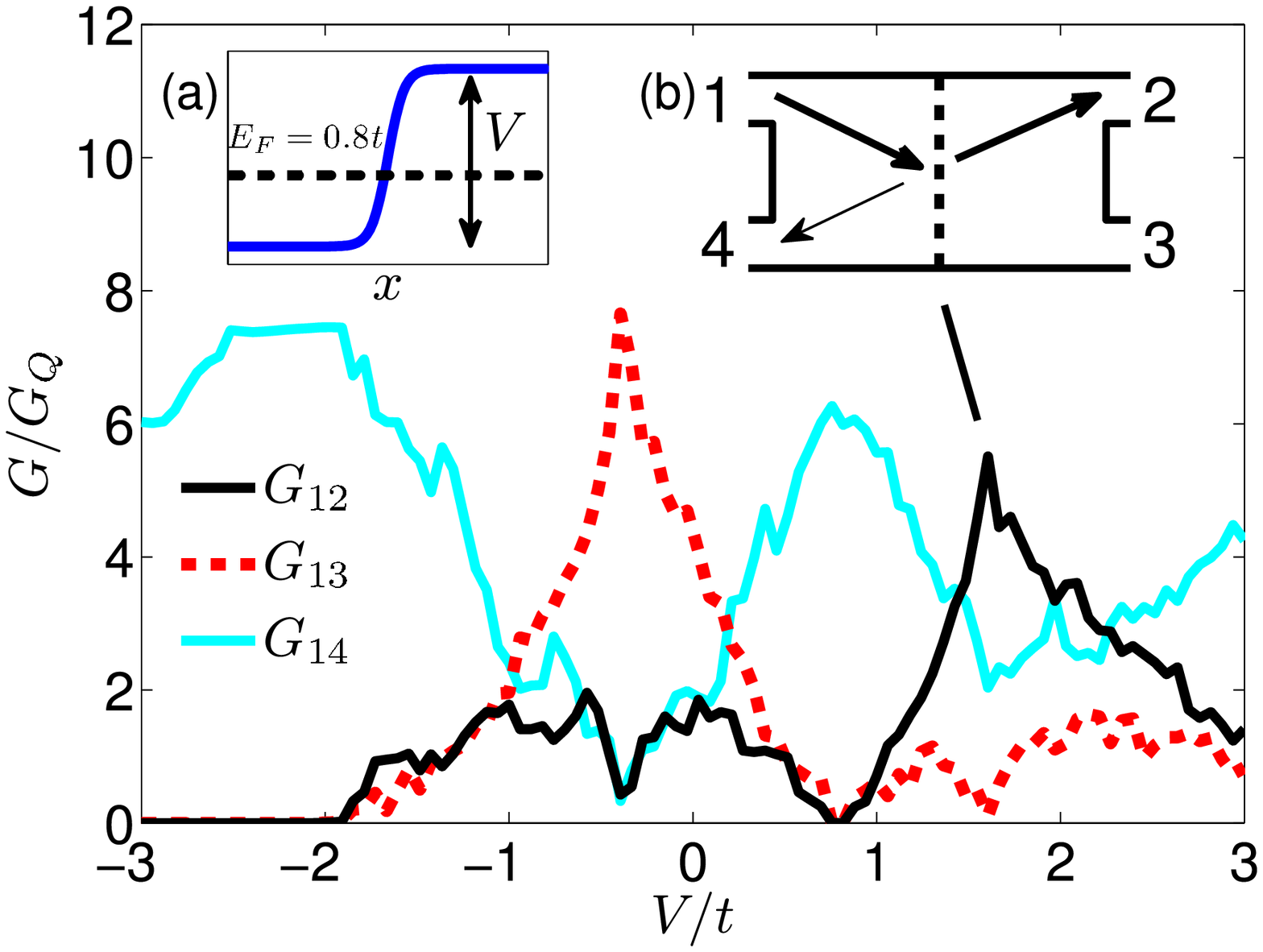}
   \caption{(Color online) The cross conductances as a function of the step height $V$ for a steep potential barrier. The central region is of size $(104,52)$. Insets: (a) Energy diagram in the longitudinal direction. (b) Schematic representation of the current flow for $V=2E_F$. The barrier works as a Veselago lens with refractive index $n=-1$, focusing the current to terminal $2$. Part of the current is reflected to terminal $4$.}
   \label{fig:step}
  \end{figure}

In the rest of the paper, we study how we can control the direction of the current using a potential step. The results for the cross conductances as a function of the height of the step are shown in Fig. \ref{fig:step}. The shape of the potential is $V(x)=V\left[1+\exp(-2x/d)\right]^{-1}$, with a smoothening factor $d=a$. This is a very steep barrier, and one gets practically the same results for $d \rightarrow 0$. In Fig. \ref{fig:step}, one can see three regions of primary interest: (i) For $V \lesssim 0$ the current flows mainly to terminal $3$ due to the tendency to propagate along the zigzag path. (ii) For $V \approx 0.8t$ the right part of the system is at the Dirac point and most of the current is reflected from the barrier to terminal $4$. The evanescent modes do not contribute here, since also the semi-infinite leads in terminals $2$ and $3$ are at the Dirac point. (iii) For $ V \approx 2E_F$ the barrier works as a Veselago lens with a refractive index $n=-1$, and the current transmits to terminal $2$. This is the central result of our paper: the current can be switched between two terminals using a pn-junction.

To understand the difference between points (i) and (iii), let us briefly review the principle of Veselago lensing \cite{Veselago1968}. Consider a case in which the left half of the graphene island is at chemical potential $\mu_L$ and the right half at $\mu_R=-\mu_L$, meaning that there is a symmetric np-junction in the middle of the island. Incoming electrons with a propagation angle $\theta$ can tunnel through the barrier as holes. The propagation angle of the outgoing hole is then $\theta'=-\theta$, since the conservation of transverse momentum requires $k_c \sin \theta = -k_v \sin \theta'$, where $k_c$ and $k_v$ are the Fermi wave vectors in the $n$ and $p$ regions, respectively. Thus the barrier works as a Veselago lens with a refractive index $n=-1$, meaning that we can focus the charge carriers to terminal $2$ instead of terminal $3$. Note that the maximum of $G_{12}$ is roughly $3/4$ of the maximum of $G_{13}$, indicating an approximate probability of $3/4$ for transmission through the np-junction. This is in reasonable agreement with the theoretical value of $T=\cos^2 \left(30^{\circ} \right) = 3/4$.

Increasing the smoothness of the barrier reduces the cross conductance $G_{12}$, since theoretically, transmission through a symmetric, smooth pn-junction is given by $T=\exp \left(-\pi k_F \tilde{d} \sin^2 \theta \right)$, where $\tilde{d}$ is the effective length scale of the step and $k_F$ is the Fermi wave vector \cite{Cheianov2006}. But in this case, the current flows to terminal $3$ for $V=0$ and reflects to terminal $4$ for $V=2E_F$, meaning that the switching effect is not lost but modified.

It is worth mentioning that we have considered a symmetric structure, in which the number of atoms in the leads in the transverse direction is odd. Due to the small size of the computational domain, the effects of breaking this symmetry are significant. The precise form of the leads is not, however, critical to our results: one can reproduce qualitatively similar results in the entire energy range using square lattice leads, when the local potential in the leads is chosen to correspond to a good description of contacts \cite{Schomerus2007}.


In conclusion, we have studied cross conductances in a four-probe all-graphene system. It was shown that in our setup aspect ratios of $\tan^{-1} (W/L)=30^{\circ}$ lead to strong turning currents due to the preference of current to propagate along a zigzag path, when the Fermi level is high. Once this situation is realized, the current can be controlled by introducing a potential step in the middle of the island. Smoothening the potential destroys Veselago lensing to some degree, but even in this case one can achieve switching between two terminals. The fact that the leads were connected to zigzag edges is important, and we leave studies on different kinds of setups for future work.

\begin{acknowledgments}
  We thank all members of the NOKIA/TKK graphene collaboration for
  helpful discussions.
\end{acknowledgments}

\end{document}